\input epsf
\documentstyle[twoside,fleqn,espcrc2]{article}

\newcommand{\grs}{$\gamma$-rays \,}

\newcommand{\AmS}{{\protect\the\textfont2
                  A\kern-.1667em\lower.5ex\hbox{M}\kern-.125emS}}

\def\Cl{$C_{L}$}

\title{The HEGRA Experiment \\
        Status and Recent Results}

\author{F.A. Aharonian\address{Max-Planck-Institut f\"ur Kernphysik,
                                 P.O.BOX 103980,
                                  D-69029 Heidelberg, Germany }
and
G. Heinzelmann \address{Universit\"at Hamburg, II. Inst. f\"ur
              Exp. Physik, Luruper Chaussee 149, 
              D-22761 Hamburg, Germany}
for the HEGRA collaboration 
}

\begin{document}
\begin{abstract}

\bigskip 
\noindent
A report is given on the status and recent results from the HEGRA
imaging atmospheric Cherenkov telescopes (IACT) concerning Mrk 501 and
GRS 1915+105 and on recent results from the HEGRA arrays concerning
galactic and extragalactic $\gamma$-sources, counterparts of TeV
Gamma-Ray-Bursts, cosmic ray anisotropies, energy spectrum and
composition of primary cosmic rays. The report summarizes the
presentations given at the 15$^{th}$ European Cosmic Ray Symposium
1996.
\end{abstract}

\maketitle

\section{GENERAL}

\begin{figure}[t]
 \vspace*{-0.6cm}
 \setlength{\epsfxsize}{7.1cm}
 \setlength{\epsfysize}{9.1cm}
 \mbox {\epsffile{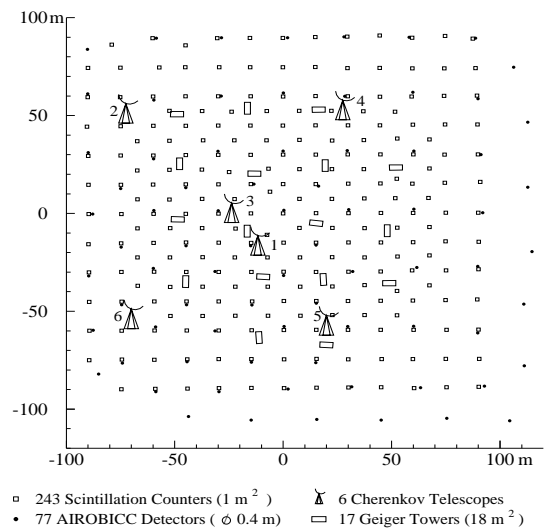}}
 \vspace*{-1.0cm}
 \caption{\label{layout} Layout of the HEGRA experiment}
 \vspace*{-0.4cm}
 
 \end{figure}

\noindent
The HEGRA (High Energy Gamma Ray Astronomy) experiment is located at
the Observatorio del Roque de los Muchachos on the Canary Island of La
Palma, (2200 m a.s.l., 28.75$^{0}$ N, 17.89$^{0}$ W) covering an area
of about 200m $\cdot $ 200m. The layout is shown in Fig.\ref{layout}.
The experiment consists of 6 imaging air Cherenkov telescopes and 3
arrays, the scintillator array, the wide angle Cherenkov array
AIROBICC \cite{AIRO} and the Geiger tower array\cite{GEIG}.

One of the major aims of the HEGRA experiment is the search for
galactic and extragalactic sources of TeV $\gamma$-radiation and a
detailed study of their characteristics with the imaging atmospheric
Cherenkov technique at energies above 500 GeV.  Hereby the
stereoscopic observation with the system of telescopes will be
exploited. The TeV $\gamma$-astronomy is continued towards higher
energies ($\geq$ 20 TeV) with the three arrays.  For the arrays a
further important aim is a measurement of the energy spectrum and
elemental composition of primary cosmic rays in the energy region from
$\approx$ 500 TeV to $\approx$ 10 PeV.

In the following first the status and performance of the imaging
atmospheric Cherenkov telescopes and their recent results concerning
Mrk 501 and the microquasar GRS1915+105 are given, then the recent
results from the arrays are presented concerning galactic and
extragalactic $\gamma$ sources, TeV counterparts of GRB's, cosmic ray
anisotropies and the energy spectrum and elemental composition of the
primary cosmic rays.

\newpage
\section{THE HEGRA SYSTEM OF IACTs}

\vspace{3mm}

\noindent
\subsection{Telescopes}

\vspace{3mm}

\noindent
Presently we are very close to complete a stereoscopic system
consisting of 5 imaging telescopes (reported by G.Hermann and M.Ulrich
at this symposium) which during the next several years will serve as
one of the most powerful instruments in the field, operating at
energies above 500~GeV \cite{Calg93} .

\begin{figure}[t]
\epsfxsize=7.5 cm
\epsfysize=5.5 cm
\epsffile[0 0 240 180]{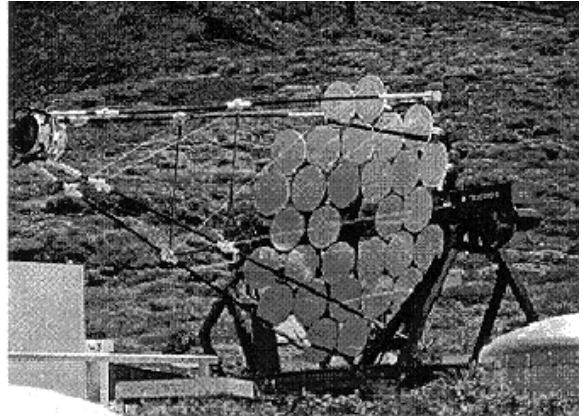}
\caption{\label{telescop} \protect \small The HEGRA telescope CT-3.}
\end{figure}

The multi mirror reflector of each telescope (Fig. \ref{telescop}) has
an area $S_{\rm mir} \simeq 8.5 \, \rm m^2$. The 60~cm diameter front
aluminized and quartz coated spherical glass mirrors with a focal
length of about 5~m are independently mounted on an almost spherical
frame of an alt-azimuth mount.  The angular resolution of the
reflector is better than 10 \mbox{arcminutes.} The mount is driven by
two stepping motors, and controlled by shaft encoders which provide a
tracking accuracy of $0.02^{\circ}$.

\begin{figure}[t]
\epsfxsize=7.5 cm
\epsfysize=5.5 cm
\epsffile[0 0 362 244]{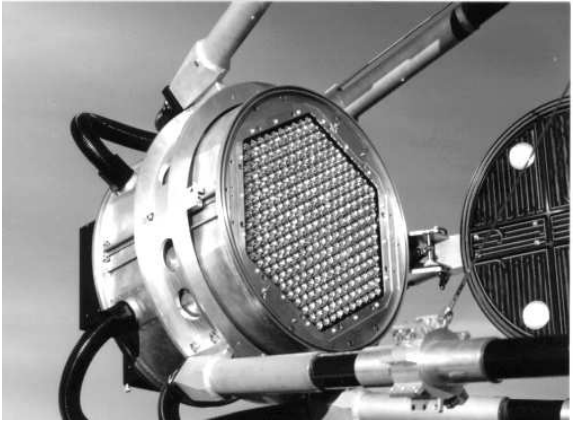}
\caption{\label{camera} \protect \small The 271-channel camera.}
\end{figure}

Each telescope is equipped with a 271-channel camera \cite{Panter95}
consisting of a close-packed cluster of circular PMTs (EMI~9083 KAFL)
faced with hollow hexagonal-to-round light guides (Fig~\ref{camera}).
The diameter of a pixel is about $0.24^{\circ}$, which results in the
telescope's field of view FoV$\approx 4.6^{\circ}$.  The photon to
photoelectron conversion efficiency including all kinds of losses (due
to the mirror reflectivity, the quantum efficiency of PMTs, losses in
light guides, {\it etc.}), and averaged over the spectrum of the
Cherenkov light at the level of observations, is around $\eta_{\rm ph
  \rightarrow e}=0.15$.

The entire electronics, including the trigger, readout, and the
telescope control, is based on the VME standard.  The digitization of
the PMT pulses is performed with a Flash ADC system which allows to
reconstruct the shape and time distribution of the pulses.  The
trigger decision is taken on two levels \cite{German95}.  The fast
trigger requires a minimum number of pixels to be fired, and the
second (slower) level restricts the topology of an image. Near each
telescope there is a small hut which houses the local electronics.
Each telescope is equipped with a local CPU which communicates via an
Ethernet line with the central computer.  Although each telescope is
built as stand alone detector, all telescopes can be integrated into
the system by software command through the central computer (Pentium
PC).

The first nominal `system-telescope' called (for chronological
reasons) CT-3, is in operation since August 1995. The data taken in
the 95/96 season generally confirm the expected performance of a
single telescope.  The next 3 telescopes CT-4, CT-5, and CT-6 have
been installed, tested and calibrated during July-December 1996, and
are now operating together with CT-3 as a system in the 2-fold
coincidence mode. The preliminary results are very encouraging and
already demonstrate the power of the stereoscopic approach (see
below).  Currently, we use different FADC systems in different
telescopes, but in the near future (until March 1997) we plan to
operate the above four telescopes with the 120~MHz FADC system. At
approximately the same time we plan to replace the present 61-channel
(pixel size $0.43^{\circ}$, and FoV$\simeq 3.9^{\circ}$) camera of
CT-2 by a standard 271-pixel high resolution camera.  Thus we hope
that after the test and calibration work, we will be able to run in
the second half of 1997 the whole system as a very sensitive
instrument for deep study of VHE $\gamma$-rays in the energy region
0.5-10~TeV.

In addition we plan to upgrade the first (prototype) HEGRA telescope
CT-1.  This equatorial mount telescope with $5 \, \rm m^2$ reflector
was installed and is taking data since 1992 \cite{CT1}. In 1994 the
original 37-channel camera was replaced by a new high resolution
127-channel camera (pixel size $0.24^{\circ}$) based on the same
pixels used in the `system' telescopes, but designed in the CAMAC
standard \cite{CT_Hires}. In 1997 the round glass mirrors will be
replaced by light hexagonal aluminum mirrors with a total reflector
area $\simeq 12 \, \rm m^2$. Thus the energy threshold of the
telescope will be reduced to 500~GeV. We plan to use this telescope
for individual observational programs, especially in immediate
reaction to potential VHE $\gamma$-ray emitters showing flaring
activity in other wavelengths. At the same time this telescope being
located very near to the central telescope CT-3, can be effectively
integrated, if necessary, with other IACTs in order to set up 2 or 3
independent stereoscopic IACT systems for simultaneous observations of
different sources.

%

\vspace{3mm}
\noindent
\subsection{Performance of the IACT system}
\vspace{3mm}

\begin{figure}[t]
\epsfxsize=6 cm
\epsfysize=6 cm
\epsffile[0 20 265 265]{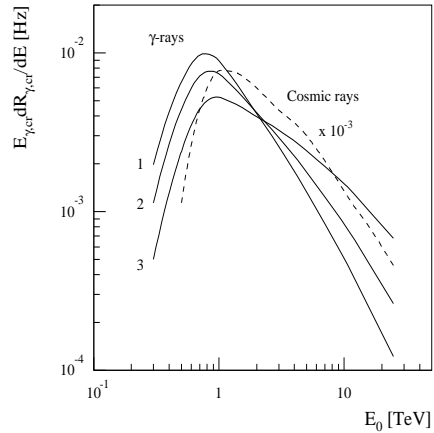}
\caption{\label{rates} \protect \small The hardware 
  differential detection rates of cosmic rays and $\gamma$-rays by
  CT-3. The rates for $\gamma$-rayes are calculated assuming a
  power-law spectrum with spectral indices: (1) -- $\alpha$=3; (2) --
  $\alpha$=2.5; (3) -- $\alpha$=2, and for an absolute flux of
  $10^{-11} \, \rm ph/cm^2 s$ above 1~TeV.}
\end{figure}

\noindent Although $\gamma$-ray showers of energy $500 \, \rm GeV$
with core position within 100~m from the telescope produce $\geq 100$
photoelectrons in the camera, sufficient for proper image analysis,
the energy threshold of a single telescope is somewhat larger, and is
determined by the suppression efficiency of the night sky background
(N.S.B.)  at the trigger level. The hardware trigger criterion for
each telescope, optimized by Monte Carlo calculations, is organized
such that the signals in at least 2 neighboring pixels from the inner
169 pixels should exceed $\sim 15$ photoelectrons ($2NN/169 \geq 15$).
The Monte Carlo calculations of the detection rates of $\gamma$-rays
from a point source, presented in Fig. \ref{rates}, show that the
effective energy threshold of the telescope CT-3 is between 700~GeV
and 1~TeV, depending on the slope of the primary photon spectrum. The
first observations with CT-3 generally confirm the calculated
performance of a single telescope, namely, a detection rate of
$\gamma$-rays from the Crab Nebula above 700~GeV of about $1.5 \times
10^{-3} \, \rm photons/s$ on top of a cosmic ray (CR) detection rate
$\approx 5 \, \rm Hz$, the significance of the signal per 1 hour of
observations being $\simeq 2.2 \sigma$.

%
\begin{figure}[t]
\epsfxsize=6.5 cm
\epsfysize=7. cm
\epsffile[80 160 550 640]{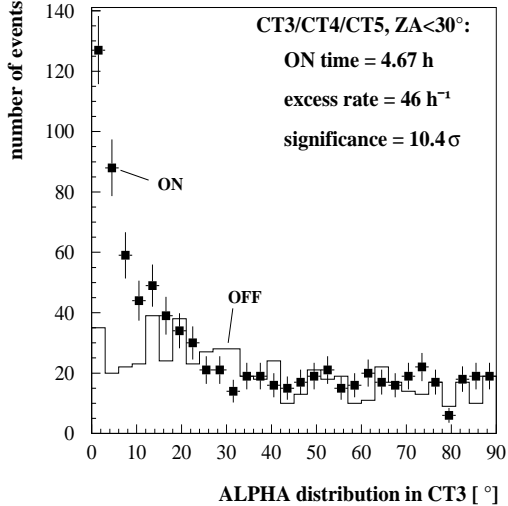}
\caption{\protect \label{system} \small ALPHA-distribution for the Crab 
in the camera of CT-3 after the application of `standard' 
image cuts in $\geq 2$ telescopes.}
\end{figure}
\begin{figure}
\epsfxsize=8 cm
\epsfysize=4 cm
\epsffile[0 0  600 305]{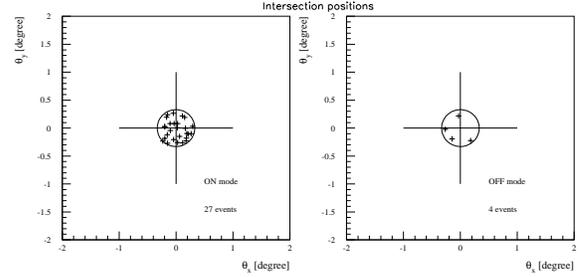}
\caption{\protect \label{stereo} \small The reconstructed coordinates of
the coincidence events observed by CT2 and CT3 within $0.3^{\circ}$
of the direction to the  Crab (ON source events) and offset of the Crab
(OFF source events) for $8.5$ hour observations in both cases.}  
\end{figure}

The requirement of simultaneous detection of showers by locally
triggered $\geq 2$ telescopes significantly reduces the effect of the
N.S.B. as well as that of the another important background connected
with local muons.  This allows a lower energy threshold of the system
down to $\leq 500 \, \rm GeV$. Also, the ability of the system to
analyze the Cherenkov images in different projections where the images
are only partially correlated, provides a superior rejection of
hadronic showers. The preliminary results of observations of the Crab
Nebula in October~'96 by CT3/CT4/CT5, operating in the 2-fold
telescope coincidence mode, generally confirm this prediction.  In
Fig. \ref{system} we present the ON-source and OFF-source ALPHA
distributions for the Crab in the camera of \mbox{CT-3}, obtained
after filtering the events through the `standard' image cuts of the
showers detected by $\geq 2$ telescopes. This preliminary result shows
that the significance of detection of $\simeq 500 \, \rm GeV$
$\gamma$-rays from the Crab is achieved at a level of $\simeq 5
\sigma$ per 1 hour.  In fact, we expect further significant
improvement of the signal using {\it energy} and {\it distance}
dependent `cuts' to the detected images.  (Note that the stereoscopic
approach allows to determine the energy and the distance to the shower
core with accuracies $\leq 20 \%$ and $\sim 10 \, \rm m$,
respectively).  Furthermore, the obvious advantage of the joint
operation of $\geq 2$ telescopes is the possibility to determine,
without any {\it a priory} assumption, the arrival directions of {\it
  individual} $\gamma$-ray primaries with accuracy $\leq 0.1^{\circ}$.
For the case of point-like sources, this provides by itself, rejection
of the CR background based only on the orientational characteristics
of showers by a factor of at least several 100.  The procedure
obviously requires good understanding of the characteristics of all
telescopes linked in the system.  Although the work regarding the
CT3/CT4/CT5 data set is still in progress, our results obtained
earlier using the combination of telescopes CT1/CT2 \cite{Kohnle}, and
telescopes CT2/CT3 allows us an optimistic view concerning the
practical realization of this important aspect of the stereoscopic
approach. Indeed, despite the short time of simultaneous observations
from the Crab by CT-2 and CT-3 taken during the season 1995/1996 and a
low coincidence rate (due to the high energy threshold of CT-2), the
stereoscopic approach reveals a clear signal with 23 `excess' events
against 4 (!) `background' events within an angle $0.3^{\circ}$ around
the direction to the Crab (see Fig.  \ref{stereo}), the latter being
in good agreement with the Monte Carlo predictions for the angular
resolution of the CT2/CT3 system.

Thus, our current understanding of the system based on the recent
experimental results as well as on comprehensive Monte Carlo studies,
allows us to make conclusive predictions for the performance of the
system of the 5 HEGRA imaging telescopes, namely

\noindent
$\bullet$ energy threshold around 500~GeV;

\noindent
$\bullet$ angular and energy resolutions of about $0.1^{\circ}$ and 
$\leq 20 \%$, respectively;

\noindent
$\bullet$ minimum detectable fluxes of $\gamma$-rays 
$F_{\rm min} \sim 10^{-12} \, \rm ph/cm^2 s$ from 
point sources, and $F_{\rm min} \sim 10^{-11} \, \rm ph/cm^2 s$ 
from moderately
extended sources with an angular size $\sim 1^{\circ}$. 

\vspace{3 mm}

\subsection{The recent results} 
\vspace{3 mm}

\noindent
While the 5-IACT system is nearing the final stage of its commission,
we continue to use the available telescopes in different
configurations not only for methodological purposes, but also for
observations of a number of astronomical objects. The first positive
signal of $\gamma$-rays was seen from the Crab Nebula in 1992 by CT-1
in its first version \cite{CT1}. Later we have detected TeV \grs from
the Crab with a significance of $\sim 10 \sigma$ using CT-2
\cite{CT2_Crab}. The current sensitivity of the system consisting of
CT3/CT4/CT5 has achieved a level which allows to obtain a strong
signal from the Crab for just a few hours of observations under almost
background-free conditions (see Fig. \ref{system}).  Thus, we can now
use the high energy photon beam of this persistent source for
calibration purposes and for an understanding of detector
characteristics.

Over the last several years CT-1 and CT-2 have been used for the
search for TeV $\gamma$-rays from other objects, in particular Cygnus
X-3, AE Aquary, {\it etc.}, reported earlier by different groups as
possible episodic $\gamma$-ray emitters (see {\it e.g.}
\cite{Weekes92}).  Unfortunately, we did not find DC $\gamma$-ray
signals from any one of these sources. However, the observations of
Markarian~421 in 1995 resulted in the independent detection of
statistically significant positive signals by CT-1 and CT-2
\cite{Hegra_Mrk421}, thus confirming the discovery by the Whipple
collaboration of TeV emission from this BL Lac object
\cite{Whipple_Mrk421}.

\begin{table}
\caption{\label{mrktab} Summary of Mrk~501 and Crab observations with CT-1.}
\begin{center}
\begin{tabular}{llllll} \hline
object                            & Mrk~501  & Crab     \\ \hline
ON-source obs. time [h]           & 146.8     & 22.9    \\
ON events after all cuts          & 1325      & 423      \\
expected background               & 974       & 237      \\
number of excess events           & 351       & 186       \\
excess rate [$\rm h^{-1}$]        & 2.4       & 8.1       \\
stat. significance (Li \& Ma)     & 5.2       & 7.6       \\ \hline
\end{tabular}
\end{center}
\end{table}
%
\begin{figure}[t]
\epsfxsize=6 cm
\epsfysize=7 cm
\epsffile[35 0 270 500]{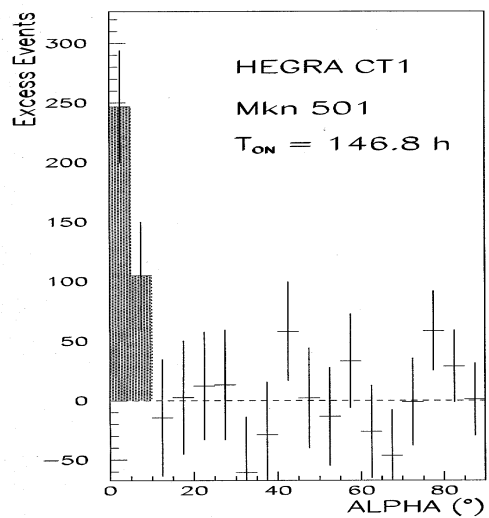}
\caption{\label{mrk501}\protect \small The ALPHA-distribution of the excess
events for Mrk~501.}
\end{figure}
\vspace{3mm}

In 1996 we allocated an essential part of the time of the available
telescopes for observations of relatively nearby AGNs, in particular
Mrk~421 (CT1, CT2, CT3), Mrk~501, PKS 2209, MS 01166 (CT1), W Comae
(CT1, CT2, and CT3, in the framework of the multiwavelength campaign),
and 1ES2344+514 (CT3/CT4/CT5).  Also, CT2 and CT3 were used for
observations of the superluminal galactic source GRS~1915+105, and
CT3/CT4/CT5 for observations of the shell type supernova remnant
$\gamma$ Cygni.  The analysis of most of these data is still in
progress.  Here we briefly discuss the results obtained from Mrk~501
and GRS~1915+105.
 
\vspace{3mm}

\centerline{{\it Mrk~501 (reported by S. Bradbury)}}  

\vspace{3mm}

This source is very similar to Mrk~421, and represents the same
sub-population of AGNs called BL Lac objects - highly variable AGNs
without strong emission lines, but showing a strong nonthermal
component of radiation from radio to X-ray wavelengths. Recent EGRET
observations showed that these objects are effective $\gamma$-ray
emitters at energies $\geq 100 \, \rm MeV$ \cite{AGN_EGRET}.  Although
Mrk~501 is not in the list of $\gamma$-ray emitting BL Lac's detected
by EGRET, this source has been reported by the Whipple group to
radiate in $\gamma$-rays above 300~GeV at an average flux level of
approximately 0.1~Crab \cite{Whipple_Mrk501}.  The low flux of VHE
$\gamma$-radiation from this source makes its detection rather
difficult for our CT-1 and CT-2 telescopes. However, at the beginning
of this year Mrk~501 started to show a flaring activity at VHE
$\gamma$-rays (Weekes, private communication). This motivated us to
allocate the CT-1 for long-term observations of Mrk~501.  In order to
obtain maximum exposure time the source was observed in tracking mode.
Almost 150~hours of good quality data was obtained at zenith angles
$\leq 25^{\circ}$ during the period March-August 1996.  The OFF-source
observations required for background determination were made when
Mrk~501 was not observable.

\begin{figure}
\epsfxsize=8 cm
\epsfysize=4.3 cm
\epsffile[50 300 500 500]{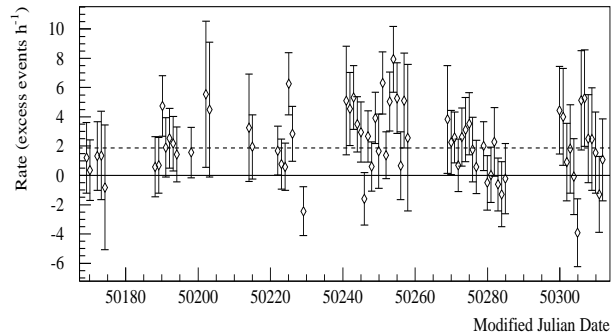}
\caption{\label{mrk501d} \protect \small Daily excess $\gamma$-ray event rates from
Mrk~501.} 
\end{figure}
\vspace{3mm}

The ALPHA distribution for Mrk~501 shown in Fig. \ref{mrk501}
indicates an excess of 351 events against expected $974 \pm 57$
background events (see Table \ref{mrktab}) which can be interpreted as
detection of $\gamma$-rays with a rate $R_\gamma \simeq 2.4 \, \rm
h^{-1}$ and statistical significance $5.2 \sigma$ \cite{HEGRA_Mrk501}.
Using the same procedure we find an excess rate of $\gamma$-like
events from the direction of the Crab, $R_\gamma \simeq 8.1 \, \rm
h^{-1}$. This implies that during the observation period the flux from
Mrk~501 was $\approx 3$ times lower than the Crab flux, i.e.
approximately $2.5 \times 10^{-12} \, \rm ph/cm^2 s$.  For comparison,
the Whipple observations of Mrk~501 in 1995 showed lower flux relative
to the Crab ($\approx 10 \%$) at energies $\geq 300 \, \rm GeV$.  This
can be explained by different energy spectra of the Crab and Mrk~501
and/or by a strong variability of Mrk~501 which can be significant on
timescales of 1~day \cite{Whipple_Mrk501}.  The daily excess
$\gamma$-ray event rates of our data are shown in Fig. \ref{mrk501d}.
Above our threshold of 1.5 TeV we have seen no significant "burst
behaviour".  Although the analysis of the distribution of total light
content in the camera does not show a significant deviation of the
spectral shape of Mrk~501 from the Crab spectrum,
due to the  uncertainty in the estimated power-law index $2.6 \pm 0.5$
we cannot make strong statements about modifications to the 
power law spectrum such as cut offs.

\vspace{3mm}

\centerline{\it GRS~1915+105  (reported by F.A.   Aharonian)}

\vspace{3mm}

The recent discovery of the galactic superluminal source GRS~1915+105
\cite{Mirabel}, the scaled down analog of the AGN jets which are
recognized as high energy $\gamma$-ray emitters, motivated the HEGRA
collaboration to put this ``microquasar'' into the ``Targets of
Opportunity'' list.

\begin{figure}[t]
\epsfxsize=6 cm
\epsfysize=7 cm
\epsffile[0 0 320 400]{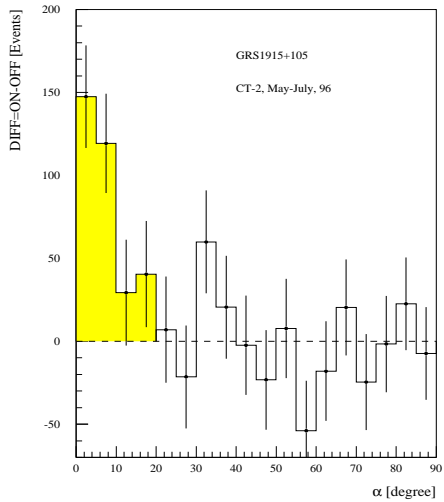}
\caption{\label{grs} \protect \small The ALPHA-distribution of the excess
events for GRS~1915+105.} 
\end{figure}
\vspace{3mm}

\begin{table}
\caption{\label{grstab} Summary of GRS 1915+105 observations with CT-2.}
\begin{center}
\begin{tabular}{llllll} \hline
                                & May  & June  & July  & Total \\ \hline
ON-source obs. \\ 
time  [h]                  & 9.89 & 37.12 & 23.83 & 70.84 \\
ON events after \\
all cuts                   & 427  & 1527  & 944   & 2898  \\
expected \\
background                 & 365  & 1373  & 875   & 2612  \\
number of \\
excess events              & 62   & 154   & 69    & 285   \\
excess rate [$\rm h^{-1}$] & 6.2  & 4.2   & 2.9   & 4.02  \\
stat. significance \\
(Li \& Ma)                 & 2.68 & 3.46  & 1.95  & 4.66  \\ \hline
\end{tabular}
\end{center}
\end{table}

The first observations of the source in Spring 1995 with the
telescopes CT1 (13~h) and CT2 (25~h) in ON/OFF mode showed a small
evidence for a signal with statistical significance $2.8 \sigma$ and
$2.6 \sigma$, respectively. However, the observations by the same
telescopes after Aug 15 with doubled exposition time did not reveal
any indication of a signal.  Since this might be a result of a highly
sporadic behavior of GRS~1915+105, we decided to continue observations
in 1996.

In 1996 GRS~1915+105 has been observed  
during the period May-July 1996 by CT2 and CT3.

CT2 with an energy threshold $\approx 1 \, \rm TeV$ had been
successfully used earlier for observations of the Crab \cite{CT2_Crab}
and Mrk~421 \cite{Hegra_Mrk421}.  The main characteristics of this
telescope are studied by Monte Carlo simulations which reproduce quite
well the distributions of different image parameters.  The rejection
power of the telescope based on both {\it shape} and {\it orientation}
parameters allows to suppress the CR background by a factor of 100,
the $\gamma$-ray acceptance being at the level of $\geq 50 \%$.  The
sensitivity of the telescope based on the observations of the Crab
Nebula is estimated as $\approx 1.25 \sigma /\rm h$ detection rate of
a signal from a point source at the flux level $J_\gamma(\geq 1 \, \rm
TeV) \approx 10^{-11} \, \rm ph/cm^2 s$.  Remarkably, this parameter
has not been changed noticeably during the last 2 years which
demonstrates the high stability of operation of the telescope.

\begin{figure}[th]
\epsfxsize=8 cm
\epsfysize=4.3 cm
\epsffile[5 20 500 240]{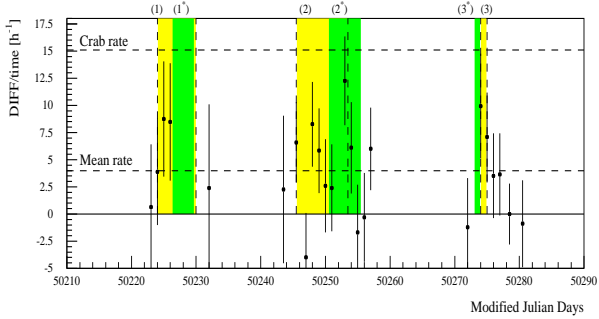}
\caption{\label{grsd} \protect \small Daily rates of the excess events
  from GRS 1915+105. For comparison, the periods of large outbursts in
  radio and X-rays also are shown. The light hatched zones (1,2,3)
  with borders indicated by dashed vertical lines corresponds to the
  periods of X-ray flares reported by Greiner J. et al., (Ap.J.
  Letters (in press); IAU Circ.6435), and the dark hatched zones
  ($1^*,2^*,3^*$) correspond to the radio flares detected by the Ryle
  and Nancay telescopes (IAU Circ. 6411; 6432).  A strong hard X-ray
  outburst in May 24-27 (MJD 50228-50231) has been reported also by
  the BATSE team (IAU Circ.6411). Note that at the time intervals
  without $\gamma$-ray data points correspond to the moonlight periods
  when the source could not be monitored.} \vspace*{0.5cm}
\end{figure}

Since GRS~1915+105 is a highly variable object, we observed the source
in the tracking mode. In addition, each month certain ON/OFF pairs
were taken in order to monitor the shape of the background
``$\alpha$-distribution''.  The statistics of observations at zenith
angles $\theta \leq 30^{\circ}$ and the results are summarized in
Table \ref{grstab}.  The ALPHA-distribution of ON-OFF source events
shown in Fig. \ref{grs} indicate the existence of a possible signal
with significance $4.7 \sigma$.  The average rate of excess events is
$\approx 4 \, \rm h^{-1}$ which corresponds to $\approx 25 \%$ of the
$\gamma$-ray detection from the Crab.  Note that most of the
``excess'' events are due to the June period of observations which can
be explained by $\approx 50 \%$ of statistics contained in this
period.  At the same time, perhaps one cannot exclude an intrinsic
variability of the signal itself.  For example the statistical
significance of the signal in June 19 alone is of about $3 \sigma$,
with a rate exceeding the average rate by a factor of 3.  This is seen
in Fig. \ref{grsd} where we present the daily rates of the ``excess''
events. Another feature of the time profile of the rate is its rather
interesting coincidence with 3 prominent few-day-duration flares of
GRS~1915+105 detected in radio and X-rays (see Fig. \ref{grsd}).  This
possible correlation in time of the rate of ``excess'' events may
serve as an additional argument that strengthens the belief that the
TeV $\gamma$-ray detection from this source is a real effect rather
than a result of statistical fluctuation of the background.  If so,
assuming that the radiation originates in the jet of GRS~1915+105
approaching us, the intrinsic average TeV luminosity of the source is
estimated as $10^{36} \, \rm erg/s$.

\vspace{3mm}

However, we would like to emphasize that before going too far in the
astrophysical interpretation of this statistically marginal signal, we
obviously need independent confirmations of this result by other
instruments and groups. As was mentioned before, the source was
observed simultaneously also by the CT-3 telescope.  At the Perpignan
Symposium, a positive signal was also shown from this telescope.
Further analysis of these data however has revealed evidence for
hardware problems during this observation period in some of the camera
pixels, resulting in large spurious signals, which may fake
$\gamma$-ray topologies. Work is in progress to correct these problems
in software and to reprocess the data. At this point, no statements
concerning GRS~1915+105 signals in CT-3 data can be made.
\newpage
\section{RESULTS FROM THE ARRAYS}

\subsection{Angular resolution of AIROBICC}
\vspace{3mm}

\noindent In an analysis using Monte Carlo and experimental data it
has been shown that the angular resolution of the AIROBICC array above
a threshold of 18 TeV for all events inside the array border is
$\sigma_{63\%}$ = 0.29 $^{\circ}$. It decreases to 0.1 $^{\circ}$ at
higher energies. The corresponding angular resolution of the
scintillator array above 20 TeV is $\sigma_{63\%}$ = 0.9$^{\circ}$ and
reaches at higher energies 0.4 $^{\circ}$ \cite{SANG}. The numbers
given refer to hadron-initiated showers of vertical incidence.  For
photon-initiated showers an angular resolution of $\sigma_{63\%}$ =
0.22 $^{\circ}$ is expected according to Monte Carlo calculations for
the AIROBICC array.

\subsection{Separation of  gamma- and hadron-initiated  showers with pearl}

\smallskip 
\noindent A method to achieve a gamma-hadron separation has been
presented by J.  Prahl. The separation of gamma- and hadron-initiated
showers is, besides a good angular resolution, crucial for a search of
$\gamma$-ray sources.

The method called pearl uses first the well know fact that
electromagnetic showers decline after the shower maximum faster than
hadronic showers.  Therefore the ratio ${\rm N_{e}}$/\Cl \, of
particles measured at ground $N_{e}$ and the number of Cherenkov
photons \Cl \, is a well known separation quantity \cite{LESS}.

The other newly introduced quantity in the method pearl is related to
the larger transverse extent of hadronic showers as compared to
electromagnetic showers due to the hadronic transverse momentum
arising in hadronic interactions. The quantity used for the transverse
spread is defined by
  $ r_P :=\sum_{i} \|\vec r_i  - \vec r_C \|\,\, n_i$ / $\sum_{i}  n_i $
whereby
  $\vec r_C$ is the reconstructed core position,
  $\vec r_i$ the position of the $\mbox{i}^{th}$ scintillator counter  and
  $n_i$ the number of particles in the $\mbox{i}^{th}$ scintillator counter.
The separation quantity $ r_P$ is related to the age parameter
from a NKG fit. 

\begin{figure}[t]
\vspace{-0.0cm}
\setlength{\epsfxsize}{5.8cm}
\setlength{\epsfysize}{5.8cm}
\mbox {\epsffile{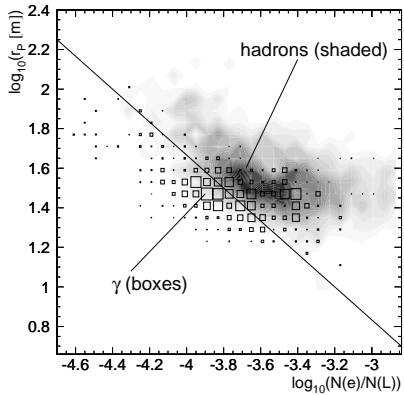}}
\vspace*{-1.2cm}
\caption{\label{pearl} \protect \small The separation
  quantities $ r_P $ vs. N(e)/N(L) (s.  text) for photon- and
  hadron-initiated showers (Monte Carlo). The straight line (pearl
  cut) indicated accepts 48 \% photon showers and rejects \mbox{90 \%}
  hadron showers (E $\geq$ 18 TeV). }

\vspace*{-0.3cm}
\end{figure}

Instead of using quantities obtained from a fit to the lateral
distributions the method pearl uses the corresponding quantities
derived from counting the amplitudes in the different counters i.e.
the sum of the pulseheights in the scintillators N(e),the sum of the
Cherenkovlight measured in the AIROBICC array N(L), and the quantity $
r_P$ sensitive to the lateral extent defined above.  The advantage of
these quantities is that showers at the trigger threshold can be
analysed without requiring a fit to converge.  (Studies have shown
that for showers for which the lateral fits converge the results are
the same in the energy range of interest).

In Fig. \ref{pearl} the two separation quantities are plotted against
each other for photon and hadron initiated showers. A clear
distinction can be seen and a straight line (pearl cut) can be used to
separate photon- and hadron initiated shower in the 2-dim logarithmic
plot.  The Q factor defined as \mbox{$ Q :=\epsilon_{\gamma} $} /
\mbox{$ \sqrt{\epsilon_{\mbox{\tiny {had}}}}$} obtained for this
method varies from Q=1.5 at threshold of 18 TeV to Q=2.5 at 50 TeV.

\subsection{Galactic TeV $\gamma$-ray sources}
\vspace{3mm}

\begin{figure}[t]
\setlength{\epsfxsize}{6.8cm}
\setlength{\epsfysize}{6.8cm}
\mbox {\epsffile{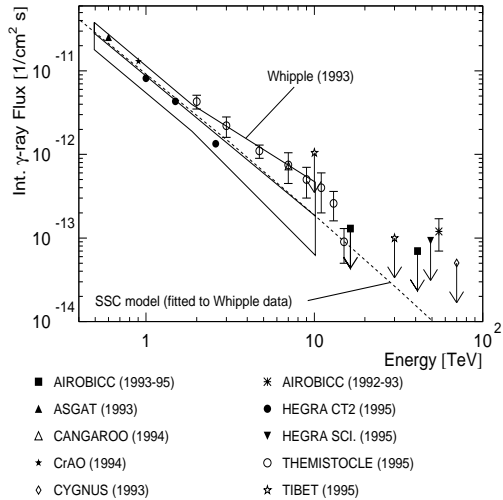}}
\caption{\label{crab} \protect \small Integral 
  $\gamma$-ray flux for the Crab nebula from the HEGRA Arrays
  (scintillator and AIROBICC) together with results from the HEGRA
  Cherenkov telescopes and other experiments.}
\end{figure}

\noindent 
With the gamma-hadron separation method pearl described above, based
on the scintillator and AIROBICC arrays, a search for galactic sources
has been performed using the 94/95 data. Amongst 13 selected sources
the most significant excess of 3.3 $\sigma$ after gamma-hadron
separation is observed for Her X-1. No excess has been observed for
the crab nebula in the 93/95 data whereas in the 1992/93 data a 3.6
$\sigma$ effect has been reported ($24^th$ ICRC Rom 1995,
unpublished).  The results for the Crab nebula are summarized in Fig.
\ref{crab}.

\subsection{ Extragalactic TeV $\gamma$-ray sources}
\vspace{3mm}
  
\noindent
Up to now two extragalactic TeV $\gamma$ sources Mrk 421 and Mrk 501
have been observed with the technique of imaging air Cherenkov
telescopes (IACT) up to $\leq$ 10 TeV.

Whereas with the IACT only one source at a time can be observed, the
arrays allow a simultaneous observation of several sources in the
field of view of HEGRA, however for the prize of a higher energy
threshold.  This may severely limit the chance of an observation due
to the absorption of TeV photons on the not well known
infrared-optical background fields ($\gamma$-ray horizon).

In order to perform a search a list of 11 source candidates with
redshifts $\leq$ 0.1 has been set up, belonging to the class of
blazars visible in the field of view of HEGRA \cite{BEAC}.  The
sources have been looked at in the stacking mode and individually.

The search has been performed for datasets with and without
gamma-hadron separation capability.  The gamma-hadron separation with
the GEIGER towers using a neural net analysis (Q-factors 2 - 2.5)
\cite{NNET} requires a threshold of 50 TeV and an angular bin of 1
degree (using the direction of the scintillator array) therefore the
results from other data sets are for comparison also evaluated for 50
TeV and 1 degree. Stacking all source candidates the highest
significances reported by N.  Magnussen are observed in the
scintillator data from 6.89 -6.92 without gamma-hadron separation (3.5
$\sigma$) and from 11.94-3.95 with gamma-hadron separation using the
Geiger towers \mbox{(4.0 $\sigma$).} The most prominent single source
candidate is 0116+319 which contributes with \mbox{4.4 $\sigma$} in
the scintillator data.  No excess has been observed for a sample of
blazars with higher redshifts and for a sample of 49 galaxies
characterized by steep radio spectra.

\begin{figure}[bh]
\setlength{\epsfxsize}{7.8cm}
\setlength{\epsfysize}{7.2cm}
\mbox {\epsffile{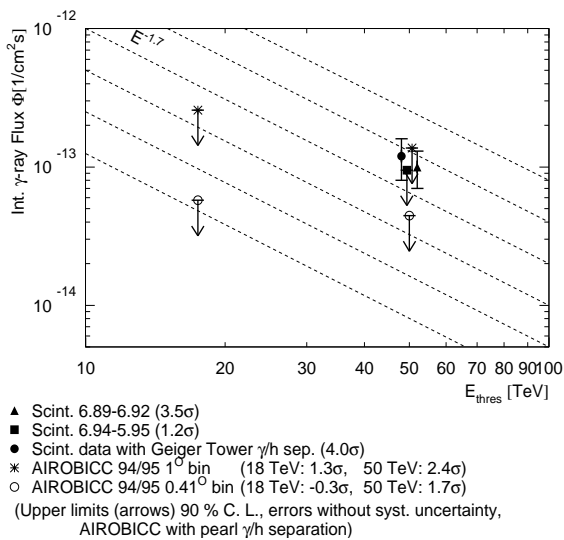}}

\vspace*{-.8cm}
\caption{\label{agn} \protect \small  Comparison of the
  results from different analyses and data samples (s. text) on a flux
  scale for a generic blazar (from a superposition of 11 nearby
  blazars)} 

\end{figure}

\newpage
 As reported by J. Prahl the scintillator data from 1994/95
show an excess of 1.2 $\sigma$ and the AIROBICC data from 1994/95 with
$\gamma$-hadron separation \mbox{2.4 $\sigma$.} Exploiting the much
better angular resolution and lower energy thresholds possible for the
AIROBICC data, the significances do not increase as expected but
decrease.  In order to compare the results of the different data sets
the excesses with the highest significance are converted tentatively
into a flux and the other results in upper limits (Fig. \ref{agn}).

A definite conclusion cannot yet be drawn and more systematic checks
are under way especially concerning the absolute pointing which is
crucial in order to exploit the good angular resolution of the
AIROBICC array. Also further AIROBICC data from 1992/93 are
investigated.  If a signal could be established this would imply, that
the infrared background radiation has to be much less than presently
assumed and the sources according to present day models would probably
be proton accelerators.

\subsection{\mbox{TeV counterparts of gamma-ray bursts}}
\vspace{0.3cm}

\noindent
Gamma-Ray-Bursts (GRB's) are known for more than 25 years yet their
physical origin and distance scale are still unknown.  Observations of
TeV counterparts would impose limits on the unknown distance scale
because of the absorption at the infrared intergalactic background
fields which limits the mean free path of TeV Photons to
distances$\leq$ 100 Mpc.

To search for TeV emission associated with gamma-ray bursts registered
by the BATSE instrument all HEGRA detector components have been used.
Especially searches with the large field of view of the air shower
arrays have been performed for TeV emission which may occur
simultaneous, preceding or following a burst.  A delayed emission may
be expected if GRB's account for the acceleration of protons up to the
highest energies which in turn generate a secondary intergalactic
electromagnetic cascade due to interactions with the microwave
background fields.

\begin{figure}[t]
\setlength{\epsfxsize}{5.7cm}
\setlength{\epsfysize}{5.cm}
\mbox {\epsffile{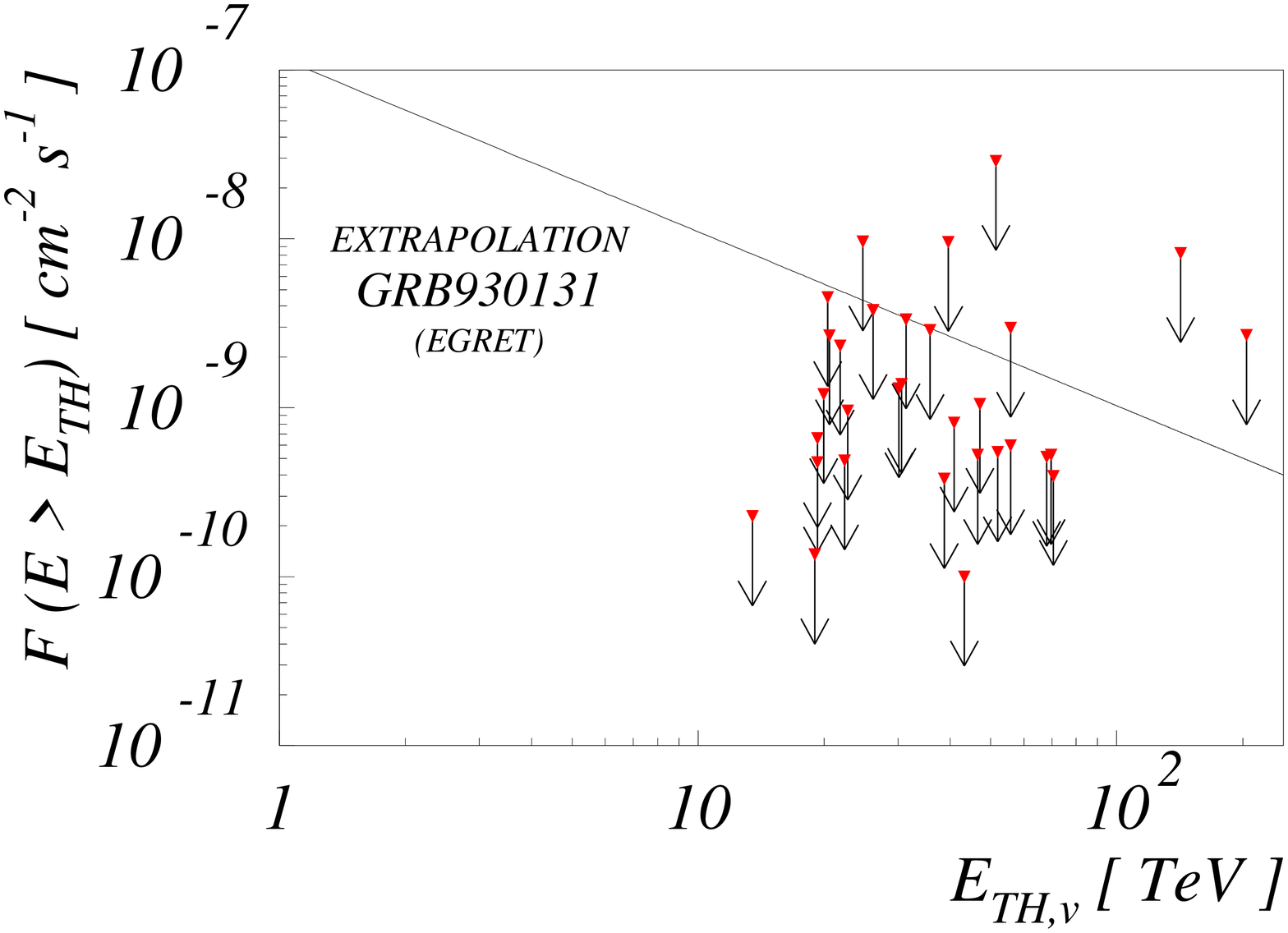}}
\vspace*{-0.9cm}
\caption{\label{grb} \protect \small For 31  BATSE bursts
  (1994-1996) the HEGRA upper limits on the coincident integral flux
  (90\% Confidence Level) as a function of the threshold energy for
  inducing photons are shown.  The threshold energy is a function of
  the zenith angle of the location of a burst. } \vspace{-0.5cm}
\end{figure}

No significant signal has been observed for the different searches as
reported by B.  Funk and H. Krawczynski.  The results from a search
for coincident emission obtained for 31 BATSE bursts are shown in Fig.
\ref{grb}.  To guide the eye an extrapolated GRB spectrum \cite{SOMM}
is displayed.

In future a lower energy threshold of about 2 TeV will be possible due
to higher data taking rate and an angle sensitive trigger.

\subsection{Cosmic ray anisotropies}
\vspace{0.3cm}

\noindent 
A search for anisotropies has been reported by D. Schmele using the
data taken during 1994/95 with the scintillator array (60 Mio events
), with the AIROBICC array (50 Mio) and in the data with the AIROBICC
array using the $\gamma$-hadron separation method pearl.

In order to search for anisotropies, all data (for each data set
separately) have been used to determine a two dimensional shape of the
background distribution in local coordinates $\phi $ and $\theta$.
This shape of the background distribution is corrected for the air
pressure by a multiplicative term depending on $\theta$ and air
pressure. Using this (pressure corrected) background shape in local
coordinates a background map in equatorial coordinates is generated
according to the actual event rates recorded. In order to obtain the
actual event rates in this procedure, the experimental event rates
have first been corrected for the (variable) deadtime and air
pressure. In this way the actual up time and performance of the array
is taken into account.

The deviations of the data from the background (for bin sizes
corresponding to the angular resolution) follow a normal distribution
N(0,1) rather well. For binsizes from $1^{\circ}$ to $50^{\circ}$ an
upper limit of $\Phi_{CR}/\Phi_{BG} \leq 10^{-2}$ for the smallest bin
and $ \leq 10^{-3}$ for the largest bin is obtained. These values are
essentially determined by the statistics of $\sim$ 50Mio events per
event sample.

\begin{figure}[t]
\epsfxsize=8.0cm
\mbox {\epsffile{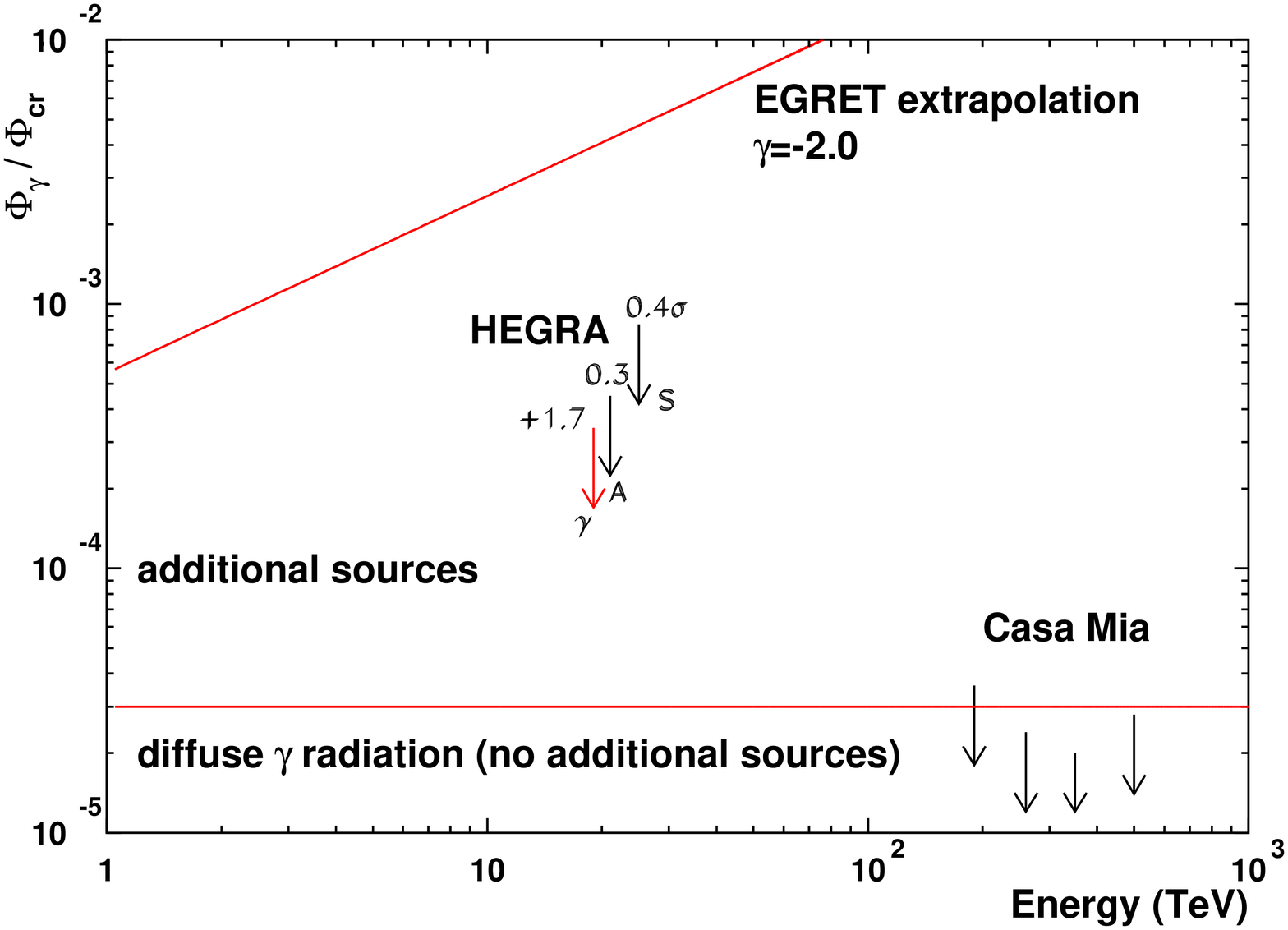}}
\vspace*{-1.0cm}
\caption{\label{disk} \protect \small $\Phi_{\gamma}/\Phi_{CR}$
  upper limit (90\% CL) from the galactic disc (S=scintillator,
  A=AIROBICC, $\gamma$= with pearl $ \gamma$-hadron separation).
  Numbers are Li-Ma significances.}
\vspace*{-3mm}
\end{figure}

In order to look for an emission from the galactic disk, the maps are
transformed into galactic coordinates and projected onto galactic
latitudes and compared.  A disk of width $ \| b \| \leq 10^{\circ}$
has been assumed (l = 30$^{o}$ - 220$^{o}$). The results for the upper
limits for a flux from the galactic disk for the three data sets are
shown in Fig.\ref{disk}. Also shown are the limits of CASA MIA
\cite{CASA}, the expected theoretical flux of $3 \times 10^{-5}$ for
diffuse emission from interaction of cosmic rays with interstellar
matter $ \| b \| \leq 5^{\circ}$ \cite{AHAD} and a naive EGRET
extrapolation with a constant slope of $\gamma=2$.  The HEGRA limits
of $\Phi_{\gamma}/\Phi_{CR} \geq 4 \times 10^{-4}$ can be interpreted
as limits for an additional flux due to 'unresolved' galactic sources
in the energy range above 20 TeV.


Finally a search for a contribution from the region around the
direction of the EHE events as seen by Fly's Eye and Yakutsk has been
performed. A contribution in the TeV energy range may be expected via
the intergalactic cascade ($p + \gamma_{3K} \rightarrow X +
\gamma_{TeV}$). On the scale from $5^{\circ}$ to $20^{\circ}$ radius
around the Fly's Eye event position (RA=85.2, DEC=48.0) \cite{FEYE},
no excess has been found. An upper limit for the contribution of an
EHE source $\Phi_{source}/\Phi_{BG} \sim 10^{-3}$ above 20 TeV can be
given.  This limit includes a possible contribution from the galactic
disk, which is near to the estimated source position.

All searches are carried out with the three data samples described
above and all three agree well within their different sensitivities
and statistics. In the near future the size of the data sample will
will be increased substantially.

\subsection{ Primary energy and composition}
\vspace{0.3cm}

\noindent 
The measurement of the energy and composition of primary cosmic rays
in the energy range of the knee is only accessible to the indirect
methods of air showers and is still controversial and a matter of
debate. Here a new method is presented to reconstruct the energy and
masses of primary particles with energies above 300 TeV from the
electromagnetic component of an air shower as measured by a
scintillator array and an array of Cherenkov counters.  The main steps
of the reconstruction procedure as presented by H. Krawczynski, can be
summarized as follows:
        
(1) The {\sl distance to the shower maximum ${ D_{max}}$ } is
reconstructed from the lateral distribution of the Cherenkov light as
measured at detection level. It can be shown by Monte Carlo simulation
that ${\rm D_{max}}$ can be determined independent of the primary mass
(and almost independent of energy). This is due to the fact that the
shower development behind the shower maximum, which dominates the
determination of the distance to the maximum, is independent of the
primary mass of nuclei.

(2) {\sl Electromagnetic energy $ E_{em}$. } From the number of
electrons $N_{e}$ as obtained from a NKG fit and the distance to the
showermaximum $ D_{max}$ the number of electrons in the maximum
$N_{e,max}$ can be inferred. This again is independent of the the
particle mass.  The number of electrons in the maximum is in a good
approximation proportional to the electromagnetic energy $ E_{em}$.
        
(3) {\sl Energy per nucleon E/A.} From the distance to the
showermaximum the penetration depth is obtained. Since the penetration
depth depends (logarithmically) on E/A, the energy per nucleon E/A can
be (coarsely) determined.

\begin{figure}[t]
    \setlength{\epsfxsize}{6.4cm} \setlength{\epsfysize}{4.5cm}
\begin{center}
    \mbox {\epsffile{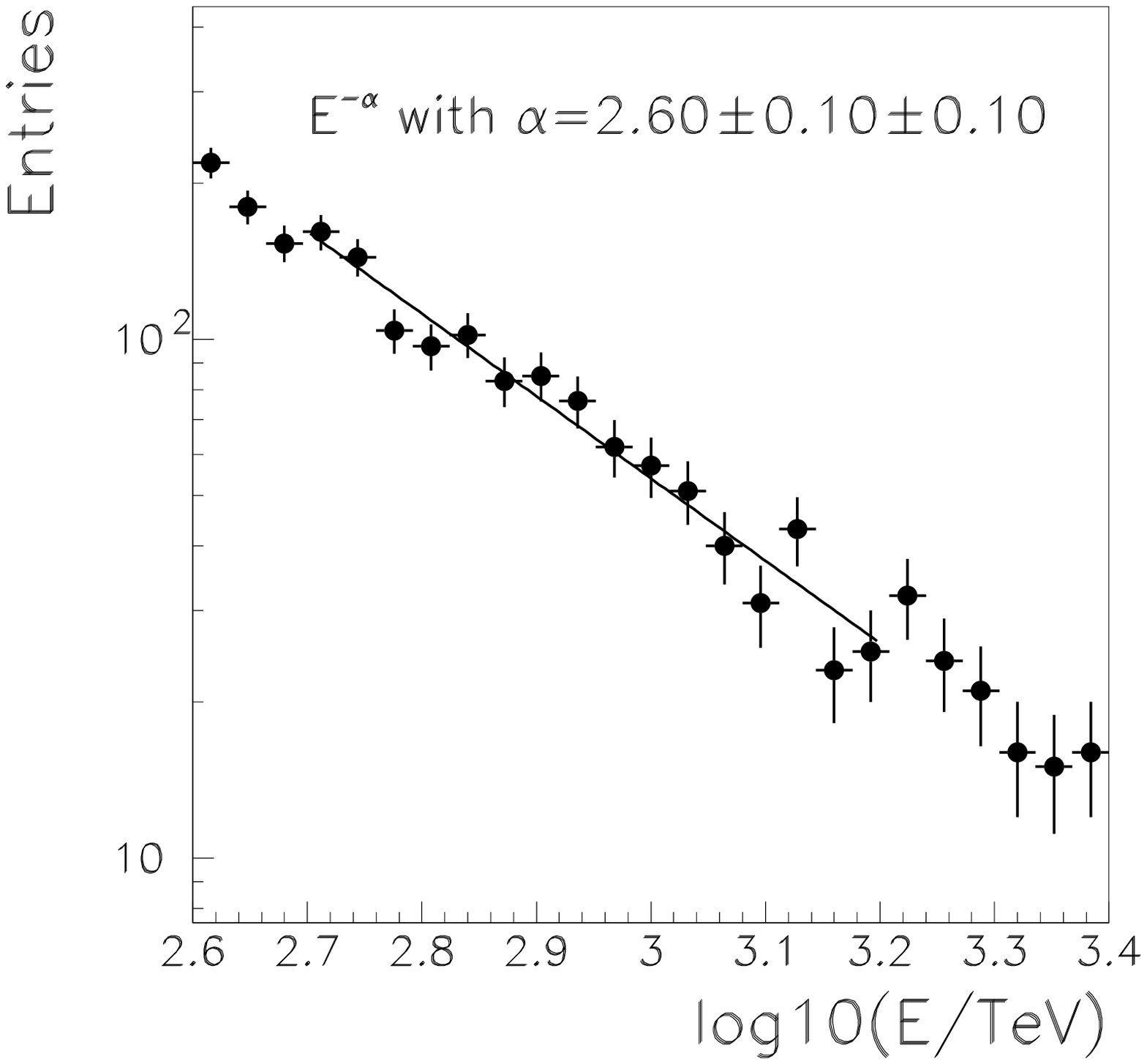}}
\end{center}
    \setlength{\epsfxsize}{5.5cm}
    \setlength{\epsfysize}{5.06cm}
\begin{center}
           \mbox {\epsffile{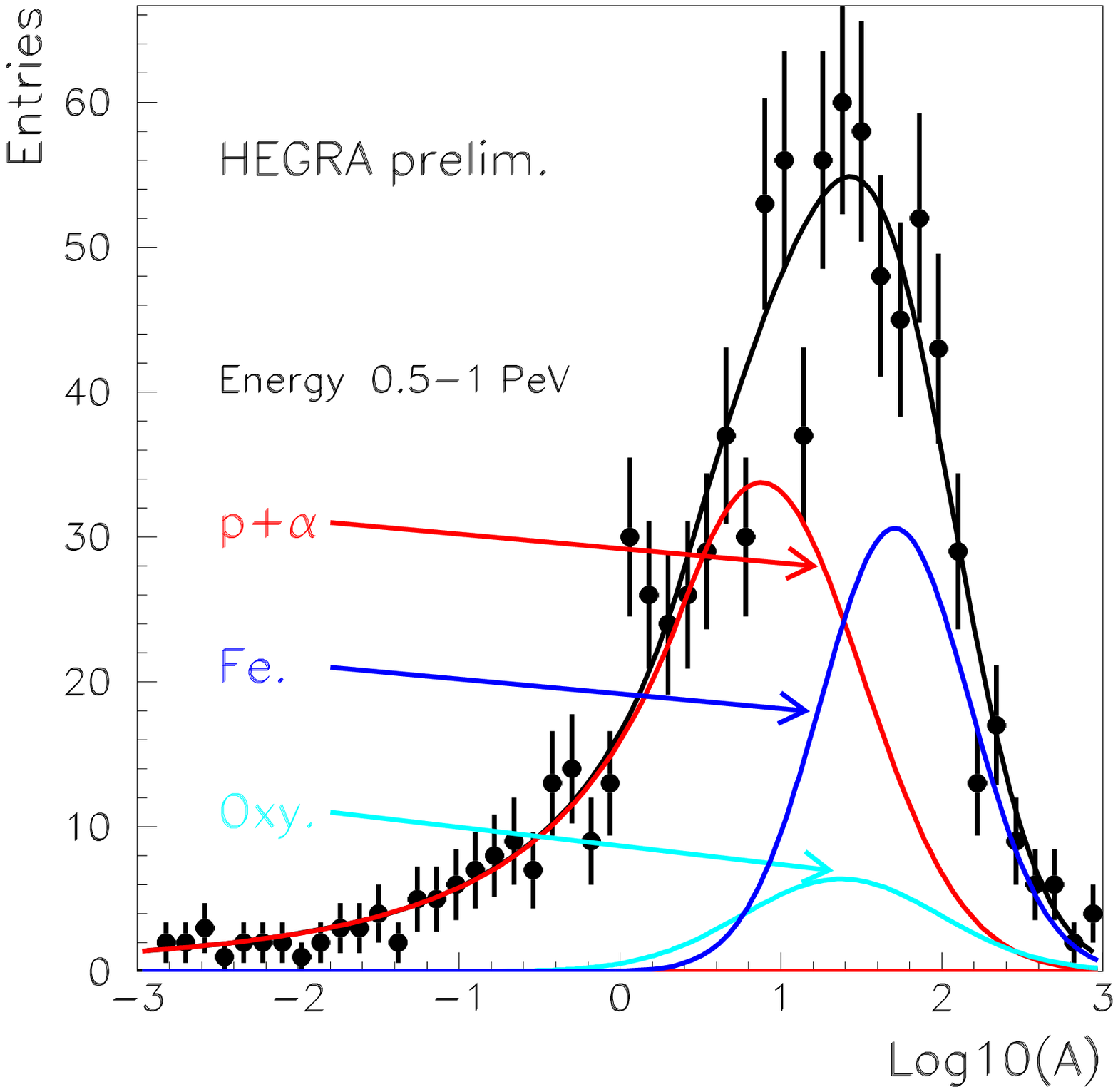}}
\end{center}
\vspace*{-1.5cm}
\caption{\label{compos} \protect \small
  The upper part shows the distribution of the reconstructed energies.
  On the lower part the distribution of the reconstructed nucleon
  number of all events with reconstructed energies between 500TeV and
  1PeV is shown. The abundances of three groups of primaries have been
  determined with a three component fit of the distributions predicted
  from Monte Carlo data for each group (dotted lines) to the measured
  distribution (solid line).  }
\end{figure}

(4) {\sl Primary energy E.} The fraction of the total energy which is
transformed into electromagnetic energy increases in a known way (i.e.
from Monte Carlo simulations) with the energy/nuleon:
\mbox{E$_{em}$=f(E/A)}.  With E/A from step (3) and E$_{em}$ from step
(2) the total energy E can be determined.

(5) {\sl Primary nucleon number A.} With E and E/A known from step (3) and (4)
also the nucleon number A is  coarsely known.
More details of the method can be found in \cite{AXEL}.

\medskip

In order to first test the new method, it is applied to the energy
range from 500 TeV to 1 PeV where the composition is known or can be
extrapolated from direct measurements taken with satellites and
balloone-borne experiment.  The preliminary results shown in Fig.
\ref{compos} are in good agreement with the results from direct
measurements. For the differential energy spectrum a spectral index of
2.60$\pm$0.10$\pm$0.1 is obtained in the energy range from 500 TeV to
2 PeV.  The abundance of light nuclei to all nuclei is obtained to be
0.62$\pm$0.05$\pm$0.10 in the energy range from 500 TeV to 1 PeV.  The
distributions in Fig.\ref{compos} reflect the fluctuations in the
penetration depth for the different primaries.  It should be stressed
that for the mass determination the mean values as well as width (i.e.
fluctuations) of the curves contribute.

Results with a slightly different method have been reported at
$24^{th}$ ICRC Rom \cite{PLAG}. In the near future the new method is
applied to the energy range above 1 PeV (knee).

\section*{Acknowledgements}

\noindent
The HEGRA Collaboration thanks the Instituto de Astrofisica de
Canarias for the use of the HEGRA site at the Roque de los Muchachos
and its facilities.  This work was supported by the BMBF, the DFG and
the CICYT.

\end{document}